\documentclass[aps,prb,reprint,superscriptaddress]{revtex4-2}

\usepackage{graphicx}
\usepackage{epstopdf}
\usepackage{dcolumn}
\usepackage{bm}
\usepackage{selinput}
\usepackage{url}

\begin{document}

\title{Low-energy electron diffraction with energy-invariant carrier-wave \\
wavenumber modulated by exchange-correlation interaction}
\author{John Rundgren}
\email{jru@kth.se}
\affiliation{Department of Physics, KTH Royal Institute of Technology, 10691 Stockholm, Sweden}

\author{Bo E. Sernelius}

\email{bo.e.sernelius@liu.se}
\affiliation{Department of Physics, Chemistry and Biology, Link\"{o}ping University, 58183 Link\"{o}ping, Sweden}

\author{Wolfgang Moritz}
\email{wolfgang.moritz@lrz.uni-muenchen.de}
\affiliation{Department of Earth and Environmental Sciences, Ludwig-Maximilians-University,\\  
Theresienstrasse 41, 80333 Munich, Germany}

\date{\today}

\begin{abstract}

We present low-energy electron diffraction (LEED) as elastic electron-atom scattering (EEAS)
operating in a target crystal waveguide where a Coulombic carrier wave is wavenumber
modulated by exchange-correlation (XC) interaction.
Carrier potential is designed using a KKR (Korringa-Kohn-Rostoker) muffin-tin (MT) model built on 
overlapping MT spheres implementing atoms with double degree of freedom, radius and potential level.
An XC potential is constructed using Sernelius's many-particle theory on electron self-energy.
EEAS phase shifts are derived from Dirac's differential equations,
and four recent LEED investigations are recalculated:
Cu(111)+$( 3\!\surd3\times\!\surd3 ) \mathrm{R30^\circ}$-TMB,
Ag(111)+$(4\!\times \!4 )$-O,
Ag(111)+$( 7\!\times\!\surd3 ) \mathrm{rect}$-$\mathrm{SO}_4$,
Ru(0001)+$( \surd3\!\times\!\surd3 ) \mathrm{R30^\circ}$-C.
TMB stands for 1,3,5-tris(4-mercaptophenyl)-benzene with chemical formula 
C$_{24}$H$_{15}$S$_3$. 
We report substantially improved reliability factors.

\end{abstract}


\maketitle

\section{\label{sec1}Introduction}

A common effort in surface structure investigations by LEED (low-energy electron diffraction) 
is to measure interatomic distances reliably with $0.001$ nm accuracy. 
LEED experience shows that the accuracy is determined by the elastic electron-atom scattering 
(EEAS) phase shifts injected as input to the diffraction code \cite{Sal94,Wal00,Vuo12}.
In a new study of EEAS we set up a KKR (Korringa-Kohn-Rostoker) muffin-tin (MT) potential 
with overlapping MT spheres corresponding to $N _\mathrm{ieq}$ symmetrically inequivalent
atoms in the surface structure unit cell.                                                                                     
It is clear that $N _\mathrm{ieq}$ radial overlaps give rise to $2N_\mathrm{ieq}$ degrees 
of freedom due to overlapping MT sphere radii and possible energy shifts of atomic potentials.
In contrast, the traditional KKR approach of  Mattheiss \cite{Mat64,Lou67,Bar07} fixes each
touching atom potential on the energy axis by superposition of neighbor potential tails,
making only $N_\mathrm{ieq}$ atomic energy shifts accessible to crystalline optimization.
Advantage overlap KKR crystal model.

We are going to design the new EEAS with similar terminolgy as used for electromagnetic scattering 
like teletransmission and radar \cite{Bri04}, with concepts waveguide, carrier-wave, and modulation by 
varying wavenumber, electronically transient and electromagnetically periodic, respectively.
The waveguide concept defines the flowchart (a)--(c) of the EEAS theory.
(a) The target propagating the incident electron serves as a waveguide for the signal-electron flow. 
A low-energy incident electron excerts spin and charge repulsion about herself, weakening
local nuclear screening.
Resulting additional nuclear attraction contributes negative self-energy $\Sigma$ that scatters
the signal electron by her own exchange-correlation (XC) potential.
In transmission electron microscopy (TEM, 200 keV--1  MeV) the signal electron traverses the
crystal with speed that makes the crystal's gas-electron and nuclear interactions seem frozen
due to the finite relaxation time of the electron gas.
(b) At very high energy the signal is guided by a Coulombic mass-and-charge potential 
that is energy invariant and acts as signal carrier potential at energies above Pauli 
indistinguishability limit $\sim\! 10$ eV.
A combined  EEAS and KKR design is set up and energy invariant MT spheres are derived.
(c) With increasing incident energy $E$ the self-energy runs through a minimum about the 
plasmon threshold $\sim\! -10$ eV at $E \sim\! 20$ eV,
so as to approach zero at very high $E$ and to become negligible.
Carrier and XC potential sum up to total EEAS potential, 
and the XC interaction manifests itself as a transient modulation of the carrier wavenumber $k$.

The paper continues with the following content:                                                           
The design of a KKR overlap MT model and the construction of the carrier potential;
the derivation of an XC potential from Sernelius's signal electron self-energy \cite{Shu87}.
Together these potentials constitute the central potential in Dirac's bispinor differential equations;
we initiate their orbital solutions.
In the interstice the big Dirac bispinor satisfies a Schrödinger equation of electron wavenumber $q$
in terms of the signal energy and an energy dependent XC potential;
we recognize the difference between $q$ and carrier wavenumber $k$ as signal electron 
wavenumber modulation.
Finally, when EEAS phase shifts are derived from the Dirac equations, we apply particular numerics
for determining the phase shift accuracy.
The advantage of overlap atomic potentials relative to touching ones is confirmed by LEED surface
structure search governed by Pendry's reliability factor \cite{Pen80}.
As test cases we use four complex surface structures published recently:
Cu(111)+$( 3\!\surd3\times\!\surd3 ) \mathrm{R30^\circ}$-TMB \cite{Sir13},
Ag(111)+$(4\!\times \!4 )$-O \cite{Rei07},
Ag(111)+$( 7\!\times\!\surd3 ) \mathrm{rect}$-$\mathrm{SO}_4$ \cite{Wyr18},
Ru(0001)+$( \surd3\!\times\!\surd3 ) \mathrm{R30^\circ}$-C \cite{Hof12}, where
TMB stands for 1,3,5-tris(4-mercaptophenyl)-benzene with chemical formula 
C$_{24}$H$_{15}$S$_3$. 

\section{\label{sec2}Energy-invariant carrier-wave}

We consider the framework of a surface unit cell of $N_\mathrm{uc}$ atoms  built by 
$N_\mathrm{ieq}$ symmetrically inequivalent atoms of interatomic separation $d_i$,
MT sphere radii $R_i$, and next neighbor MT radii $R_{\mathrm{NN}i}$, $i=1,2,\dots,N_\mathrm{ieq}$.
In particular, supposing that $R_i$ and the set of $R_{\mathrm{NN}i}$ about site $i$ would fulfill 
the equality $R_i + R_{\mathrm{NN}i} = d_i $, they would form a prototype of touching atoms.  
Quantitatively we now define MT spheres overlap by the inequality                                   
  \begin{equation}\label{eq1}   
  R_i + R_{\mathrm{NN}i} \leq  d_i \times ( 1 + S_i ), {\ } S_i\ge 0 ,  {\ } i=1,2,\dots,N_{\mathrm{ieq}}
  \end{equation}
where the parameter $S_i$ of atom $i$ indicates relative overlap with respect to touching spheres.

We initiate the EEAS theory by constructing the carrier potential.                                    
Assigned to each site $i$ is an atom of electron density $\rho_i(r)$ and nucleus $Z_i$.
We utilize computer code available from National Institute of Standards and Technology \cite{Shi95,Shi97,SM1}  
for calculating Coulombic potentials $V_{\mathrm{NIST}i}(r)$ from $\rho_i(r)$ and $Z_i$ by  
Poisson integration with standard normalization, when the potential tends to zero at large radius.
Given $V_{\mathrm{NIST}i}(r)$, we wish to build a KKR crystal carrier potential
$V_i(r)$, $r\leq R_i$, $i = 1,2,\ldots,N_{\mathrm{ieq}}$, where $R_i$ signifies MT radius.
A crucial point in the modeling of a carrier potential is that the potential level $V_{\mathrm{NIST}i}(R_i)$
is arbitrary with respect to the searched crystalline potential level $V_i(R_i)$.
Differential evolution (DE) \cite{Sto97} is an appropriate method for extracting information   
from an EEAS picture with $2N_{\mathrm{ieq}}$ degrees of freedom.
We set up a spatially constant interstitial potential $V_0$
equal to the unit cell average of the  atomic potentials at MT sphere peripheries,
  \begin{equation}\label{eq2}   
  V_0 = N_{\mathrm{uc}}^{-1} {\sum}_{i=1}^{N_{\mathrm{ieq}}} N_{{\mathrm{eq}}i} V_i(R_i)
  \end{equation}
Here, a signal electron scattering argument comes in. 
Signal electron reflection at potential steps between interstitial $V_0$  and peripheral $V_i(R_i)$
would excite alien standing-wave resonances in the MT spheres.
Removal of potential steps is found to be a necessary condition on clean signal electron 
EEAS in the crystal.
The requirement corresponds to a DE fitness number based on the $N_{\mathrm{ieq}}$ 
degrees of freedom for potential shifts,
  \begin{equation}\label{eq3}   
  \epsilon =  \min\big[\max_i ( | V_i(R_i) -  V_0 | ) \big],  {\ } i=1,2,\dots,N_{\mathrm{ieq}}
  \end{equation}
The application of DE on equations (\ref{eq1})--(\ref{eq3}) starts from the work space 
$V_i(r) = V_{\mathrm{NIST}i}(r)$, $ i=1,2,\dots,N_{\mathrm{ieq}}$.
A first DE calculation creates sets $R_i$  and $V_i(r)$ with a coarce fitness number 
$\epsilon = \mathrm{O}(1)$ in units of eV.
Before a continued DE calculation we make the following substitution,
  \begin{equation}\label{eq4}   
  V_i(r) - [V_i(R_i) - V_0] \rightarrow V_i(r),   {\ } i=1,2,\dots,N_{\mathrm{ieq}} 
  \end{equation}
noting that $V_0$ remains invariant.
With one more turn DE finishes with $\epsilon = \mathrm{O}(10^{-15})$ in a computer 
of 64 bit precision.
The generated $V_i(r)$'s form exact KKR carrier potentials.

  \begin{figure} 
  \includegraphics[angle = 0, scale = 0.3]{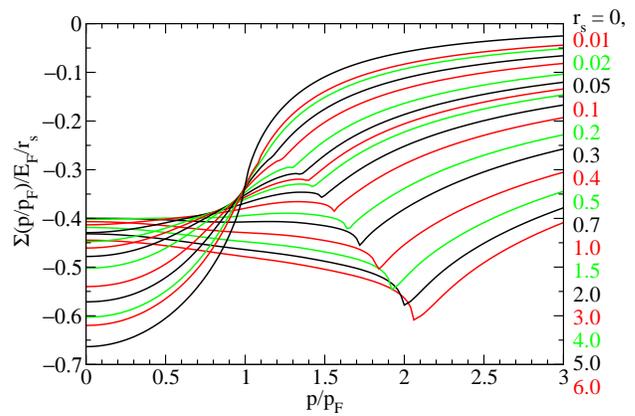}   
  \caption{\label{fig:1}  Data base 
  $\texttt{sdat}(p/p_{\mathrm{F}},r_\mathrm{s}) = 
  \Sigma(p/p_\mathrm{F},r_\mathrm{s})/E_\mathrm{F}/r_\mathrm{s}$
   in units of $a_0^{-1}$.
  The curve labeled $r_\mathrm{s}=0$ signifies the limit of  $\texttt{sdat}$ at $r_\mathrm{s} \rightarrow 0$, 
  equal to the Hartree-Fock exchange potential.
  The $\texttt{sdat}$ curves for $r_\mathrm{s}>0$ display exchange-correlation potentials 
  divided by $E_\mathrm{F}$ and $r_\mathrm{s}$;  
   kinks mark plasmon thresholds.}
  \end{figure}

The interstitial potential $V_0$ of the surface structure is now determined by connecting     
the unit cell of the surface slab to the crystal bulk.
Fortunately, a slab-to-bulk boundary condition is inherent in the DE method.
We force the MT radii of the innermost layers of the surface unit cell to optimize in minute intervals
fitted to known MT radii of the bulk;
in this way DE optimization lifts $V_0$ to bulk energy level.
In next paragraph we learn that electron self-energy is currently normalized to zero at 
TEM energy.
Correspondingly, standard carrier potential $V_i(r), \,  i=1,2,\dots,N_{\mathrm{ieq}}$ is normalized 
to $V_0 = \mathrm{MTZ}$ (muffin-tin zero) $= 0$. 

\section{\label{sec3}Electron self-energy and wavenumber modulation}

\begin{figure}    
  \includegraphics[angle = 0, scale = 0.3]{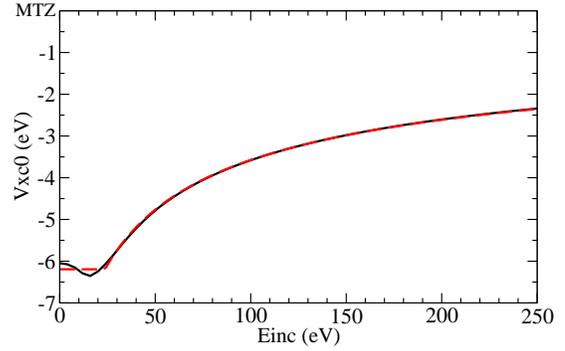}  
  \caption{\label{fig:2}
  Interstitial XC potential of surface   Cu(111)+$( 3\surd3\times\!\surd3 ) \mathrm{R30^\circ}$-TMB
  (black) and 4-coefficient approximation
  $V_{\mathrm{XC}0}(E) \approx \max \big[  p_1 +  p_2/{\sqrt{ E + p_3}}, p_4 \big]$ (red).
  It is extractable from LEED experiment by r-factor optimization of $p_1-p_4$. }
  \end{figure}

The advantage of energy dependent XC interaction in LEED investigations was published    
in 1982 \cite{Nev82}.
Hedin-Lundqvist \cite{Hed71,Wat76} had shortly before put one-electron excitation equal to 
the sum of incident energy and electron-gas chemical potential.
Later Sernelius applied Rayleigh-Schrödinger many-particle theory and identified electron self-energy
and electron gas ground-state, see Ref. \cite{Shu87} and Appendix A,
  \begin{equation}\label{eq5}    
  p^2 + \Sigma( p/p_\mathrm{F},r_\mathrm{s}) = E + E_{\mathrm{F}} +  \Sigma(1,r_\mathrm{s})
  \end{equation}
Left-hand member of eq.~(\ref{eq5}) creates kinetic energy and self-energy potential
of an excited signal electron, while right-hand member provides available energy sources, 
incident energy and chemical potential of ground-state electron gas.
Designations are:
$p=$ signal electron momentum, equal to wavenumber in atomic Rydberg units;
$\Sigma=$ signal electron self-energy, of which $\Sigma(1,r_\mathrm{s})$ is the XC part of the 
chemical potential; 
$p_{\mathrm{F}}( r_\mathrm{s})=$ Fermi momentum;  
$E_{\mathrm{F}}( r_\mathrm{s})=$ Fermi energy;
with $r=$ atomic radius, $r_\mathrm{s}(r)=$ electron-space radius, and $\rho(r)=$ electron gas density;
$(4\pi/3)\, r_\mathrm{s}^3 \rho = 1$; $r_\mathrm{s} = 0 - 6 a_0$ for most crystals.
Momentum $p/p_\mathrm{F}$ in one-electron excitation eq.~(\ref{eq5}) is solved as a function
of incident energy $E$ by means of Sernelius's data base $\texttt{sdat}$ \cite{SM2}
illustrated in Fig.~\ref{fig:1},
  \begin{eqnarray}\label{eq6}   
  (p/p_\mathrm{F})^2 = (E/E_\mathrm{F}) + 1 + r_\mathrm{s}
  [ \texttt{sdat}(1,r_\mathrm{s})\! -\! \texttt{sdat}(p/p_\mathrm{F},r_\mathrm{s}) ]                                                                     
  \nonumber\\
  \texttt{sdat}(p/p_{\mathrm{F}},r_\mathrm{s}) = \Sigma(p/p_{\mathrm{F}},r_\mathrm{s})/
  E_{\mathrm{F}}/ r_{\mathrm{s}}  \quad \quad \quad \quad
  \end{eqnarray}
Iteration of $p/p_\mathrm{F}$ is done with bivariate interpolation \cite{Abr67} in $\texttt{sdat}$.
Extrapolation of the data base $\texttt{sdat}$ to high momentum simulates the
Hartree-Fock exchange potential $\propto\!\!-(p/p_\mathrm{F})^{-2}$ \cite{Ash76} 
togther with the Lindhard correlation potential $\propto\!\!-(p/p_\mathrm{F})^{-1}$  \cite{Lin54}.
The former rapidly vanishes with energy and the latter remains finite but negligible at TEM energy.
The result is a set of atomic XC potentials $i$ in terms of variables $r$, $E$, and $p$
together with materials data $p_\mathrm{F}$, $E_\mathrm{F}$, and $r_\mathrm{s}$,
  \begin{equation}\label{eq7}    
  V_{\mathrm{XC}i}(r,E) = \texttt{sdat}(p/p_\mathrm{F},r_\mathrm{s})
  \, E_\mathrm{F} r_\mathrm{s},    {\ } i=1,2,\dots,N_{\mathrm{ieq}}
  \end{equation}
which we add up to a spatially flat interstitial XC potential,
  \begin{equation}\label{eq8}   
  V_{\mathrm{XC}0}(E) = N_{\mathrm{uc}}^{-1} {\sum}_{i=1}^{N_{\mathrm{ieq}}}   
  N_{{\mathrm{eq}}i} V_{\mathrm{XC}i}(R_i,E)
  \end{equation}
Substitution similar to expression (\ref{eq4}) makes XC potentials $ V_{\mathrm{XC}i}(R_i,E)$
equal to intersticial $V_{\mathrm{XC}0}(E)$.
Following the normalization convention of $\texttt{sdat}$ in Fig.~1 we shift the field of
XC potentials to $\mathrm{MTZ}=0$ at TEM energy.
The $V_{\mathrm{XC}0}(E)$ curve in Fig.~\ref{fig:2}  turns out to have              
universal shape given by a four-coefficient approximation \cite{Nev82,Run03},
whose increasing part almost identically fits the calculated XC potential.

In an attempt to confirm the theoretical self-energy $V_{\mathrm{XC}0}(E)$ by experimental LEED 
data, we attach a multiplier $f_\mathrm{XC}$ to the right-hand member of eq.~(\ref{eq7}).
The purpose is to consider $f_\mathrm{XC}$ as an adjustable structural parameter and to record
an r-factor versus $f_\mathrm{XC}$ curve during the LEED investigation.
$f_\mathrm{XC}$ close to unit would indicate confirmation.

\section{\label{sec4}Elastic electron-atom scattering phase shifts}

Relativistic EEAS with central potential is determined by Dirac equation                   
$W\psi = H\psi$ \cite{Ros61}, where $\psi$ is a two-component spinor and $W$ is the eigenvalue 
$ E + m c^2$ with $E =$ signal energy, $m =$ mass $\textstyle{ \frac{1}{2}}$, and
$c =$ speed of light $2/\alpha$, $\alpha$ being fine structure constant $1/137$
(atomic Rydberg units).
Eigenstates are $G_\kappa = u_{1\kappa}/r$ and $F_\kappa  = u_{2\kappa}/r$
with  spin-orbit quantum number $\kappa$, total angular momentum $j$, and atomic orbital $l$,
  \begin{equation} \label{eq9}    
  \kappa = \left \{ \begin{array}{l l l l}
   - l -1 & {\mathrm{for}} &  j=l+\frac{1}{2},  {\ }  l=0,1,2\ldots       \\
  {\ \ } l      & {\mathrm{for}}  &  j=l - \frac{1}{ 2},  {\ } l={\ \ \ }1,2,\ldots 
    \end{array} \right .
  \end{equation}
$\kappa$ is negative or positive corresponding to spin $\frac{1}{2}$ parallel or $-\frac{1}{2}$ 
antiparallel to electron's direction of flight \cite{Mot65}.
$u_{1\kappa}$ and $u_{2\kappa}$  are solutions of two Dirac differential equations with
central potential,
  \begin{eqnarray} \label{eq10}   
  \frac{du_{1\kappa}}{dr} &=&  -\frac{\kappa}{r} u_{1\kappa} + 
  \{c+c^{-1}[ E - V_{\mathrm{T}i}(r,E) ]\}u_{2\kappa} 
  \nonumber \\
  \frac{du_{2\kappa}}{dr} &=& \frac{\kappa}{r} u_{2\kappa} - 
  c^{-1} [ E -  V_{\mathrm{T}i}(r,E) ]u_{1\kappa}
  \end{eqnarray}  
where  $V_{\mathrm{T}i}(r,E) = V_i(r)+V_{\mathrm{XC}i}(r,E)$ signifies total atomic potential. 
Boundary conditions at the origin are \cite{Ros61},
  \begin{equation} \label{eq11}   
  u_{1\kappa} = A {r^{\gamma}} \quad \mathrm{and} \quad 
  u_{2\kappa} = A {r^{\gamma}} ({\kappa} + {\gamma})/{\zeta}
  \end{equation}
with $A=$ const, $\gamma^2 = \kappa^2 - \zeta^2$, $\zeta = \alpha Z$, and $Z=$ atomic number;
$\gamma >0$ corresponds to regular solutions of the Dirac equations.
We use an exponential radial grid $r_n,{\ } n=1,2,\dots$, where $r_1 = 10^{-10}$, 
a current initial radius for atomic calculations \cite{Shi95}.
With increasing orbitals $l$, angular momenta displace the electron waves 
$u_{1\mathrm{\kappa}}$, $u_{2\mathrm{\kappa}}$ away from the origin.
Correspondingly, initiation radii $r_\kappa = (r_1)^{1/{\gamma}}$ move so that wave factors 
$A (r_{\kappa})^{\gamma}$ in condition (\ref{eq11}) remain constant $A r_1$.
We initiate with $u_{1\mathrm{\kappa}} = r_1$ and 
$u_{2\mathrm{\kappa}} = r_1({\kappa} + {\gamma})/{\zeta}$, $A=1$.
The Dirac equations are then integrated using Shampine and Gordon's ODE (ordinary differential 
equation) solver \cite{Sha75a,SM3} built on Adams predictor-corrector method with automatic initiation and
step size control. 
    
In the interstice of spatially constant $V_{\mathrm{XC}0}(E)$ elimination of $u_{2\kappa}$  
from the first-order Dirac equations (\ref{eq10}) gives rise to a second-order Schrödinger 
equation for $u_{1\kappa}$,
  \begin{eqnarray} \label{eq12}   
  \frac{d^2 u_{1\kappa}}{dr^2}+ \Big [ q^2 - \frac{\kappa (\kappa + 1)}{r^2} \Big ] u_{1\kappa} = 0,
  \quad  {\ \  }
  \nonumber \\
  q^2 = E -V_{\mathrm{XC}0}(E) + c^{-2} [ E - V_{\mathrm{XC}0}(E) ]^2  
  \end{eqnarray}
The difference between wavenumbers $q$ and $k=\surd E$ constitutes the XC modulation of the carrier 
wave.
$\kappa(\kappa+1)$ equals $ l(l+1)$ for both spins, and the eigenvectors are $rj_l(qr)$ and $ry_l(qr)$ 
with spherical Bessel functions $j_l$ and $y_l$.
The signal electron in the interstice  is expressed in terms of scattering phase shifts 
$\delta_\kappa$,
  \begin{equation} \label{eq13}   
  u_{1\kappa}(r) = C[r j_l (qr) \cos\delta_\kappa - r y_l(qr) \sin\delta_\kappa ]
  \end{equation}
$C=$  const.  
$\kappa$ takes a single subscript $-1$ for orbital $l=0$ and two subscripts, $-l-1$ and $l$,
for orbitals $l \ge 1$.
The phase shifts $\delta_\kappa$ are determined by equating logarithmic derivatives belonging to
$u_{1\kappa}$ inside and outside MT radius $R$; 
$u_{2\kappa}$ of relative magnitude $c^{-1}$ is neglected.
With prime for differentiation with respect to $r$, phase shifts are obtained from
  \begin{equation} \label{eq14}   
  \tan \delta_\kappa = 
         \frac{[u _{1\kappa}(r)]^\prime rj_l(qr) -  u _{1\kappa}(r) [ rj_l(qr) ]^\prime}
               {[u _{1\kappa}(r)]^\prime ry_l(qr )-  u _{1\kappa}(r) [ ry_l(qr) ]^\prime}
          \ \Big | _{r=R_i}
  \end{equation}
The phase shift spin states define two sets of scattering amplitudes,
$q^{-1}  e^{i \delta_{-l-1}} \sin{\delta_{-l-1}}$ and 
$q^{-1}  e^{i \delta_{l}} \sin{\delta_{l}}$,
of spin $ \textstyle{\frac{1}{2}}$ and $-\textstyle{\frac{1}{2}}$, respectively.
LEED without spin combines the above scattering amplitudes with orbital weights 
$l+1$ and $l$ divided by $2l+1$.

ODE is managed by two error tolerances $\texttt{relerr}$ and $\texttt{abserr}$ and responds with  
errors $\epsilon_1 \le \texttt{relerr} {\,} |u_{1\kappa}(r^{\prime})|+ \texttt{abserr}$ 
with $r^{\prime}$ on Adams predictor-corrector radial grid.
Continued calculation transmits these errors to error bars on the EEAS phase shifts $\delta_\kappa(E)$.
Since the phase shifts provide preliminary data for reliability estimation of theoretical LEED
spectra \cite{Pen80}, we wish to work with known phase shift accuracy.
Using the benchmark technique from earlier LEED work \cite{Run07}, we accompany Dirac representation
(\ref{eq10})  by a representation that is analytically equivalent and is at once computationally different 
with respect to algorithms and initiation conditions. We apply the substitutions
  \begin{equation} \label{eq15}      
  u_{1\kappa} = y_{1\kappa} {r^{\gamma}} \quad \mathrm{and} \quad 
  u_{2\kappa} = y_{2\kappa} {r^{\gamma}} (\kappa + \gamma)/\zeta
  \end{equation}
to equations (\ref{eq10}) and find a new Dirac representation, 
  \begin{eqnarray} \label{eq16}      
  \frac{dy_{1\kappa}}{dr} &=& (\kappa + \gamma)
  \Big[ -\frac{\kappa}{r} y_{1\kappa} + 
  \{ c+c^{-1}[ E - V_{\mathrm{T}i}(r,E)] \} \frac{y_{2\kappa}}{\zeta} \Big]  
  \nonumber \\
  \frac{dy_{2\kappa}}{dr} &=& (\kappa - \gamma)
  \Big[ \frac{\kappa}{r} y_{2\kappa} -  c^{-1} [ E -   V_{\mathrm{T}i}(r,E) ] \frac{y_{1\kappa}}{\zeta} \Big]
  \end{eqnarray}
whose integration is initiated from boundary condition $ y_{1\kappa} =  y_{2\kappa}=$ unity at the origin,
distinct from initiation $u_{1\kappa} = r_1$ and $u_{2\kappa} = r_1 (\kappa + \gamma)/\zeta$, 
$r_1 = 10^{-10}$.
Equations (\ref{eq16}) integrate twice as fast as eqs. (\ref{eq10}).
The Dirac representations (\ref{eq10}) and (\ref{eq16}) of phase shifts 
$\delta_\kappa$ and $\delta_\kappa^{\prime}$, respectively, determine the phase shift accuracy
  \begin{equation} \label{eq17}   
  \epsilon = 
  \max_{i,l,E,\kappa} |{\,} ( |\delta_\kappa(E)| -   |\delta_{\kappa}^{\prime}(E)| ){\,} |  
  \end{equation}
with respect to spin, energy, orbitals, and atoms. 
For example, with $\texttt{relerr}\!=\! \texttt{1.E-6}$ and $\texttt{abserr}\!=\!\texttt{1.E-9}$ 
atoms S, C, H, and Cu of the surface structure
Cu(111)+$( 3\surd3\!\times\!\surd3 ) {\, } \mathrm{R30^\circ}$-TMB \cite{Sir13} attain phase shift 
accuracy $ \texttt{2.E-5}$.

 \begin{table}
 \caption{\label{tab:1} 
 LEED reliability \cite{Pen80} with MT spheres  nonoverlap (N) and overlap (O).
 $E$ is energy of incidence; 
 O parameter is  $S \sim [(R + R_{\mathrm{NN}})/(\mathrm{interatomic\ distance})] - 1$. }
 \begin{ruledtabular}
 \begin{tabular}{l c c c c } 
 & &\multicolumn{2}{c}{ reliability} \\
 \cline{3-4} 
 Surface structure & $E$ (eV) & N\footnotemark[1] & O\footnotemark[2] & $S$  \\
 \hline
  Cu(111)+$( 3\!\surd3\times\!\surd3 ) \mathrm{R30^\circ}$-TMB     
       & 11--200 & 0.32\footnotemark[3] & 0.252  & 0.3\\
  Ag(111)+$(4\!\times \!4 )$-O                                                    
       & 25--250 & 0.34\footnotemark[4] & 0.26 & 0.75 \\   
  Ag(111)+$( 7\!\times\!\surd3 ) \mathrm{rect}$-$\mathrm{SO}_4$   
       & 10--150 & 0.235\footnotemark[5] & 0.211  & 0.4\\   
  Ru(0001)+$( \surd3\!\times\!\surd3 ) \mathrm{R30^\circ}$-Cl         
       & 20--300 & 0.137\footnotemark[6] & 0.116  & 0.4\\  
 \end{tabular}
 \end{ruledtabular}
 \footnotemark[1]{Refs.\cite{Run03,Nas07}, nonoverlapping MT radii.}\\ 
 \footnotemark[2]{This work, overlapping MT radii.}\\
 \footnotemark[3]{Ref.\cite{Sir13}.} 
 \footnotemark[4]{Ref.\cite{Rei07}.}
 \footnotemark[5]{Ref.\cite{Wyr18}.}
 \footnotemark[6]{Ref.\cite{Hof12}.}  
 \end{table}

\section{\label{sec5} Result}

Moritz used  the LEEDFIT program packet \cite{WM20,WMVH} to implement the EEAS method 
of Sec.~II--IV for surface structure search.
Appendix B gives a short account of the successive LEEDFIT iterations determinig best r-factor
\cite{Pen80}.
Table~I illustrates LEED investigations on four complex structures carried out in earlier publications 
and in present work, using MT spheres with nonoverlap (N) and overlap (O), respectively.
LEED based on O instead of N type of KKR crystal substantially lowers the reliability factors
\cite{Sir13,Rei07,Wyr18,Hof12}: 
in value by 0.068, 0.08, 0.024, 0.021, and in percent by 21, 23, 10, 15.
The O-KKR crystal has $2N _\mathrm{ieq}$ degrees of freedom with adjustable MT radii
and MT potentials, unlike Mattheiss's N-KKR model \cite{Mat64,Lou67,Bar07} having $N _\mathrm{ieq}$ 
degrees of freedom with restricted MT radii and fixed MT potentials.
The O-KKR implementation creates potential junctions of no steps between MT spheres and interstice 
(Sec.~II); then the probability of standing-wave electron resonances in the MT spheres is negligible.
A single MT overlap parameter $S$ per structure is found to give similar reliability as atomic $S_i$.
Appendix B, Fig.~4, shows $S$ varied from zero to saturated MT overlap. 
 
For searching an experimental confirmation of the electron self-energy $V_{\mathrm{XC}0}(E)$,
Moritz included the multiplier $f_\mathrm{XC}$ in the set of structural parameters.
During $S$ variation the r-factor is recorded as a function of $f_\mathrm{XC}$ in the range 
$0.4$--$1.4$, see Appendix B, Fig.~5.
Pendry's statistical RR measure \cite{Pen80} on the LEED experiment determines the limits of trustworthy 
reliability and identifies error bars on the r-factor versus $f_\mathrm{XC}$ diagram.
The four structures in Table I are found to give self-energy multipliers gratifyingly close to unity 
\cite{Sir13,Rei07,Wyr18,Hof12}: $1.00\pm0.1$, $0.94\pm0.1$, $1.06\pm0.1$, $0.89\pm0.2$.
This is the first LEED confirmation of the electron self-energy potential.

Introduced in LEED is here the technique of energy-invariant carrier-wave and transient electron 
self-energy modulation.
It turns out that elastic electron-atom scattering in material is analogous to electromagnetic 
radar and teletranmission in space.
Electron gun and crystal target form an electron waveguide; 
the electrostatic part of the atoms in the crystal generates an energy invariant carrier-wave potential; 
and the electron gas self-energy potential excites a transient wavenumber modulation of the carrier-wave. 
Above LEED incidence $\sim\! 10$ eV the modulation is due to $\sim\! -10$  eV XC interaction, 
and beyond a few tens of eV incidence the  modulation continues as correlation interaction
$\propto\! (-1$/wavenumber) eV. 

\section*{Acknowledgment}

The work came up at the International Workshop on Quantitative LEED Analysis 
organized by Professor E.W. Plummer, Louisiana State University, April 18--20, 2016.
Financial supports from KTH Royal Institute of Technology, Ludwig-Maximilians-University,
and Link\"{o}ping University are gratefully acknowledged.

J.R. conceptualized, B.E.S. determined electron self-energy, and W.M. conducted LEED investigations.

\appendix

\section{\label{A}Electron Self-Energy Determination}
\subsection{\label{E}System of Interacting Electrons}

Our calculation of the electron self-energy is based on the Rayleigh-Schr{\"o}dinger (RS) perturbation theory or on the mass-shell-perturbation-theory. The starting point is the energy, $E$, of the system of interacting electrons. The Hamiltonian is written with a variable coupling constant, $\lambda$,
\begin{eqnarray}
H(\lambda )  &= & {H_0} + \lambda {\kern 1pt} V \nonumber \\
 H(1)  &= & H\\
 H(0)  &= & {H_0}\;, \nonumber
\end{eqnarray}
where $H_0$ the non-interacting Hamiltonian and $V$ is the interaction part, in our present case the Coulomb interaction between the electrons.

Let,
\begin{eqnarray}
H(\lambda )\left| {\left. {{\Psi _0}(\lambda )} \right\rangle } \right. &=&E(\lambda )\left| {\left. {{\Psi _0}(\lambda )} \right\rangle } \right.  \nonumber \\
\left\langle {{\Psi _0}(\lambda )} \right.\left| {\left. {{\Psi _0}(\lambda )} \right\rangle } \right. &=& 1 \\
& \Downarrow &  \nonumber \\
E(\lambda ) &=& \left\langle {\left. {{\Psi _0}(\lambda )} \right|} \right.H(\lambda )\left| {\left. {{\Psi _0}(\lambda )} \right\rangle } \right., \nonumber
\end{eqnarray}
where $\left| {\left. {{\Psi _0}(\lambda )} \right\rangle } \right.$ is the interacting ground state of the system at coupling strength $\lambda$.

The derivative with respect to the coupling constant reduces to
\begin{eqnarray}
&\frac{d}{{d{\kern 1pt} \lambda }}E(\lambda )=\left\langle {\left. {\frac{{d{\Psi _0}(\lambda )}}{{d{\kern 1pt} \lambda }}} \right|} \right.H(\lambda )\left| {\left. {{\Psi _0}(\lambda )} \right\rangle } \right. + \left\langle {\left. {{\Psi _0}(\lambda )} \right|} \right.H(\lambda )\left| {\left. {\frac{{d{\Psi _0}(\lambda )}}{{d{\kern 1pt} \lambda }}} \right\rangle } \right. \nonumber \\
&+ \left\langle {\left. {{\Psi _0}(\lambda )} \right|} \right.\frac{{dH(\lambda )}}{{d{\kern 1pt} \lambda }}\left| {\left. {{\Psi _0}(\lambda )} \right\rangle } \right.\\
&=E(\lambda )\frac{d}{{d{\kern 1pt} \lambda }}\left\langle {{\Psi _0}(\lambda )} \right.\left| {\left. {{\Psi _0}(\lambda )} \right\rangle } \right. + \left\langle {\left. {{\Psi _0}(\lambda )} \right|} \right.V\left| {\left. {{\Psi _0}(\lambda )} \right\rangle } \right. \nonumber \\
&=\left\langle {\left. {{\Psi _0}(\lambda )} \right|} \right.V\left| {\left. {{\Psi _0}(\lambda )} \right\rangle } \right..\nonumber
\end{eqnarray}

Integrating this equation with respect to the coupling constant from 0 to 1 gives
\begin{equation}
E_{int} = E - {E_0} = \int_0^1 {\frac{{d{\kern 1pt} \lambda }}{\lambda }} \left\langle {\left. {{\Psi _0}(\lambda )} \right|} \right.\lambda {\kern 1pt} V\left| {\left. {{\Psi _0}(\lambda )} \right\rangle } \right..
\label{IntE}
\end{equation}

Eq.(\ref{IntE}) is the starting point for diagrammatic perturbation theory. The result of rather cumbersome derivations is
\begin{eqnarray}
\begin{array}{l}
\!\!\!\!{E_{xc}} \!=\!  + i\int_0^1 {\!\frac{{d\lambda }}{\lambda }{\kern 1pt} } \frac{1}{2}\!\sum\limits_{\bf{q}} {'\!\!\left\{ {\int_{ - \infty }^\infty  {\frac{{d\omega }}{{2\pi }}\hbar\! \left[ {\frac{1}{{{\varepsilon ^\lambda }\left( {{\bf{q}},\omega } \right)}} \!-\!1} \right]\!\!} } \right.} 
\left. {\!-\frac{{N\lambda {v_q}}}{{iv\kappa }}} \!\right\}
\end{array}
\label{Einter}
\end{eqnarray}
where the second term in the momentum summand is a subtraction of the self-interaction contributions of the $N$ number of electrons in the system. The prime indicates that the ${\bf{q}} = {\bf{0}}$ term should be omitted. To be noted is that this expression is formally exact. The approximations lie in the approximations of the dielectric function, ${{\varepsilon ^\lambda }\left( {{\bf{q}},\omega } \right)}$. The dielectric function is the longitudinal version on time-ordered form. Please note that here we let the exchange and correlation energy, ${E_{xc}}$, denote the whole interaction energy, not the energy per electron which is usually the case.

A nice reformulation of the self-interaction in terms of the dielectric function of the system, ${\varepsilon _0^\lambda \left( {{\bf{q}},\omega } \right)}$, in case all electrons are responding completely independent of each other leads to a more symmetric expression,
\begin{eqnarray}
{E_{xc}} =  + i\int_0^1 {\frac{{d\lambda }}{\lambda }{\kern 1pt} } \frac{1}{2}\sum\limits_{\bf{q}} {'\left\{ {\int_{ - \infty }^\infty  {\frac{{d\omega }}{{2\pi }}\hbar \left[ {\frac{1}{{{\varepsilon ^\lambda }\left( {{\bf{q}},\omega } \right)}} - 1} \right]} } \right.}\nonumber \\
\left. { - \left[ {\frac{1}{{\varepsilon _0^\lambda \left( {{\bf{q}},\omega } \right)}} - 1} \right]} \right\}.
\label{Einter2}
\end{eqnarray}

We use the Random Phase Approximation (RPA) in the diagrammatic perturbation theory. Then the dielectric function is the so-called Lindhardt dielectric function \cite{Lin54} and it can be expressed in terms of the polarizability, ${\alpha _0}\left( {{\bf{q}},\omega } \right)$,
\begin{equation}
{\varepsilon ^\lambda }\left( {{\bf{q}},\omega } \right) = 1 + \lambda {\alpha _0}\left( {{\bf{q}},\omega } \right).
\end{equation}

We may perform the integration over coupling constant with the result:
\begin{equation}
{E_{xc}} =  - i\frac{1}{2}\sum\limits_{\bf{q}}{' {\int_{ - \infty }^\infty  {\frac{{d\omega }}{{2\pi }}\hbar \ln \left[ {\frac{{\varepsilon \left( {{\bf{q}},\omega } \right)}}{{{\varepsilon _0}\left( {{\bf{q}},\omega } \right)}}} \right]} }}.
\label{Einterimag}
\end{equation}

\begin{figure}
\includegraphics[angle = 0, scale = 0.5]{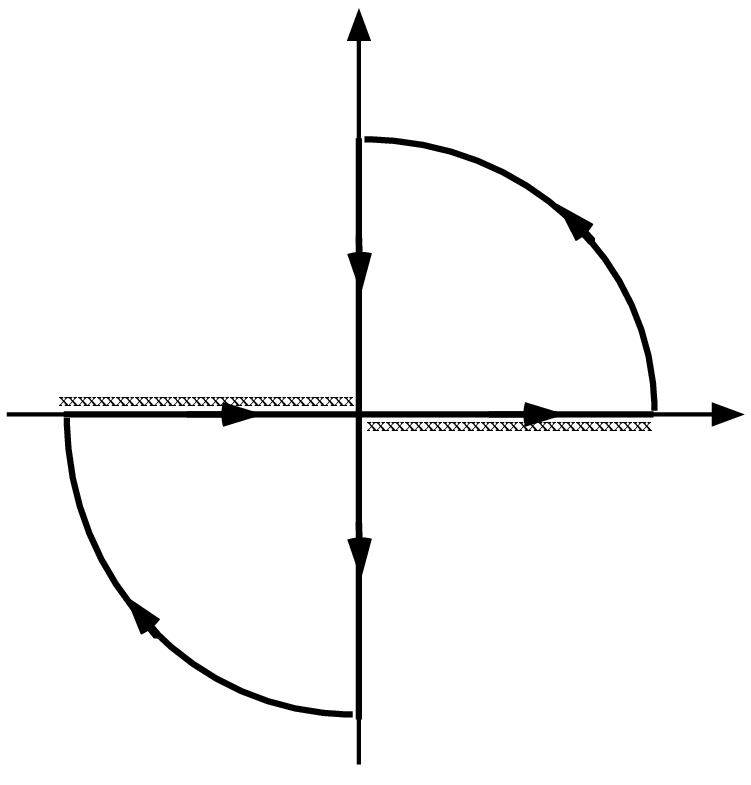}
\caption{\label{fig:3} Integration contour in the complex frequency plane. 
The crosses indicate the positions of the poles of the integrand when the functions are on the time-orderd form}
\label{contour}
\end{figure}

Now, the integrand is rather complex and complex valued. It is favourable to perform integration along the imaginary axis in the complex frequency plane. We deform the integration path along the contour in Fig.\ref{contour}. All poles are outside the contour. This means that the result of the integration is zero. Now the contributions from integration along the two curved parts of the contour cancel. This means that integration along the real axis gives the same result as integration along the imaginary axis in positive direction.

Thus,
\begin{equation}
{E_{xc}} = \frac{1}{2}\sum\limits_{\bf{q}}{' {\int_{ - \infty }^\infty  {\frac{{d\omega }}{{2\pi }}\hbar \ln \left[ {\frac{{\varepsilon \left( {{\bf{q}},i\omega } \right)}}{{{\varepsilon _0}\left( {{\bf{q}},i\omega } \right)}}} \right]} }}.
\label{Long}
\end{equation}

It is interesting to point out that exactly the same result is obtained from the summation of the zero-point energies \cite{Ser18} of the longitudinal normal modes of the system. This alternative derivation is much simpler. Here we also find contributions from transverse modes with the result
\begin{equation}
{E_t} = 2\frac{1}{2}\sum\limits_{\bf{q}}{' {\int_{ - \infty }^\infty  {\frac{{d\omega }}{{2\pi }}\hbar \ln \left[ {\frac{{{\varepsilon _T}\left( {{\bf{q}},i\omega } \right){\omega ^2} + {{\left( {cq} \right)}^2}}}{{{\omega ^2} + {{\left( {cq} \right)}^2}}}} \right]} }},
\label{Trans}
\end{equation}
where the extra factor 2 in front of the right-hand-side is due to the two polarization directions of the transverse normal modes.

These results turn out to be negligible \cite{Ser18}. For the interaction between objects the first type, that comes from longitudinal modes, is responsible for the van der Waals interactions while the other type, that comes from transverse modes, gives rise to Casimir interactions. The dielectric function, ${{\varepsilon _T}\left( {{\bf{q}},i\omega } \right)}$, in Eq.(\ref{Trans}) is the transverse version. It coincides with the longitudinal version, ${\varepsilon \left( {{\bf{q}},i\omega } \right)}$, as used in Eq.(\ref{Long}), for small $q$ but differs for large $q$. 

Why did we not obtain this contribution using the diagrammatic perturbation theory? The reason is that we did not include the whole interaction part in Eq.(\ref{Einter}), only the part representing the Coulomb interaction between the electrons. There are also ${\bf{p}} \cdot {\bf{A}}$- and ${\bf{A}} \cdot {\bf{p}}$-terms, where ${\bf{A}}$ is the vector potential.

Our derivation of the electron self-energy is based on the results in Eq.(\ref{Long}).

\subsection{\label{Self}The self-Energy}

The electron self-energy is the energy shift of a single particle state caused by interactions. All interactions cause energy shifts of varying importance. For the electron gas the shift comes from the exchange and correlation effects. If all electron states are shifted the same amount there is a rigid shift of the electron band. In that case most, but not all, of the physics is unchanged; the work function, e.g., is altered. 
	If the shifts are not exactly rigid the energy dispersion is modified and there is a deformation of the density-of-states. This leads to a change in band width and in effective mass. In most cases the change in dispersion near the chemical potential has the greatest effect on the properties of the system since most processes involve low-energy excitations. 
	What we have said so far refers to the real part of the self-energy. The self-energy is in general complex valued. The imaginary part is a measure of the life-time of a particle in that particular state. A state with a large imaginary part is short lived. The life-time is intimately connected to the mean-free-path. The imaginary part will furthermore cause a broadening of experimental spectra. 
	
In the calculation of the interaction energy of an electron gas the energy can be viewed as getting contributions from processes where electrons are scattered out of occupied states into unoccupied states.

In RS the self-energy for a state above the chemical potential is the change in the total interaction energy of the system when an electron is placed in this state. Placing the electron in that state adds new possible processes where the electron is scattered out from the state; it also reduces other processes, processes where electrons otherwise could have been scattered into the state. These processes are now forbidden due to the Pauli exclusion principle.

The self-energy for a state below the chemical potential is minus the change in the total interaction energy of the system when an electron is removed from this state. Removing the electron from that state (or adding a hole) adds new possible processes where other electrons are scattered into the state; these processes are no longer forbidden by the Pauli exclusion principle. It also reduces other processes, processes where the electron otherwise could have been scattered out of the state. 

Now, to get the energy of state $\left( {{\bf{p}},\sigma } \right)$, what one really does is to take the variational derivative of the total energy $E$ of the system with respect to the occupation number ${n_{{\bf{p}},\sigma }}$ of state $\left( {{\bf{p}},\sigma } \right)$,
\begin{equation}
{\varepsilon _{{\bf{p}},\sigma }} = \frac{{\delta E}}{{\delta {\kern 1pt} {n_{{\bf{p}},\sigma }}}} = \varepsilon _{{\bf{p}},\sigma }^0 + \hbar {\Sigma _{{\bf{p}},\sigma }}.
\label{En}
\end{equation}
Tim Rice \cite{Rice65} was the first to point this out. The self-energy is then
\begin{equation}
\hbar {\Sigma _{{\bf{p}},\sigma }} = \frac{{\delta {E_{xc}}}}{{\delta {\kern 1pt} {n_{{\bf{p}},\sigma }}}}.
\label{EQSE}
\end{equation}
In RS the self-energy of an electron in an electron gas depends on the momentum and spin of the electron and on the electron density. RS has proven to give reliable results in a number of situations. Semiconductors made metallic, either by heavy doping or high excitation, can act as very flexible model systems of metals. We will give some examples below.

Germanium is a many-valley semiconductor. Heavily $n$-doped germanium has four equivalent Fermi-volumes in the Brillouin Zone. When a stress is applied to the system some valleys move up in energy and some move down; this results in shrinking and growing Fermi-volumes. This has effects on optical and transport properties. How the valleys move is closely related to the real part of the self-energy. The most straight forward test of the real part of the self-energy is found in the optical birefringence experiment and the agreement between theory \cite{Ser83} and experiment is striking.

GaAs is a direct-band-gap semiconductor. Heavily $p$-doped GaAs can be used to produce spin-polarized electrons by using polarized light. The excited electrons in the conduction band will then recombine with the holes in the two valence bands. The comparison \cite{Ser86} between theory and experiment on the resulting luminescence polarization gives a thorough test of both the real and imaginary parts of the self-energy. The RS self-energy well passed the test.

More recent tests have been made. The imaginary part of the self-energy has successfully been tested \cite{Meng15} in the dynamics of highly photo-excited electrons and holes in silicon. 

In \cite{Ser83} one applied a mechanical stress on a heavily doped many-valley semiconductor to move the conduction band valleys up or down in energy. An alternative way is to apply a strong magnetic field. Then spin-up and spin-down valleys move up or down in energy. Fermi-volumes of one type grow and of the other type shrink, leading to a change in transport properties. Both silicon \cite{Silva15} and germanium \cite{Silva20} show negative magnetoresistance. The comparison between experiment and theory means a test of the real part of the self-energy. The results show qualitative agreement with a negative magnetoresistance in both the experimental and theoretical results. 

Let us now return to the actual calculation of the self-energy in Eq.(\ref{EQSE}).
We should note that when the change in occupation numbers has been made the system is no longer in its ground state; it is unstable and decays; the signature of an unstable system is a complex-valued energy.

The occupation numbers are found in the polarizabilities entering the expression for the
energies. Now,
\begin{eqnarray}
\begin{array}{l}
{\alpha _0}\left( {{\bf{q}},\omega } \right) =  - \frac{1}{\hbar }{v_q}\sum\limits_\sigma  {\int {\frac{{{d^3}k}}{{{{\left( {2\pi } \right)}^3}}}} } \\
\int\limits_{ - \infty }^\infty  {\frac{{d\varepsilon }}{{2\pi {\kern 1pt} i}}} G_\sigma ^{\left( 0 \right)}\left( {{\bf{k}},\varepsilon } \right)G_\sigma ^{\left( 0 \right)}\left( {{\bf{k}} + {\bf{q}},\varepsilon  + \omega } \right),
\end{array}
\label{alfa}
\end{eqnarray}
where
\begin{eqnarray}
\begin{array}{l}
G_\sigma ^{\left( 0 \right)}\left( {{\bf{k}},\varepsilon } \right){\kern 1pt}  = \frac{{{n_{{\bf{k}},\sigma }}}}{{\varepsilon  - {{\hbar {k^2}} \mathord{\left/
 {\vphantom {{\hbar {k^2}} {2{m^*}}}} \right.
 \kern-\nulldelimiterspace} {2{m^*}}} - i\eta }} + \frac{{1 - {n_{{\bf{k}},\sigma }}}}{{\varepsilon  - {{\hbar {k^2}} \mathord{\left/
 {\vphantom {{\hbar {k^2}} {2{m^*}}}} \right.
 \kern-\nulldelimiterspace} {2{m^*}}} + i\eta }};\;\\
{n_{{\bf{k}},\sigma }} = \left\{ \begin{array}{l}
1\;{\rm{if}}\;k < {k_{\rm{F}}}\\
0\quad {\rm{otherwise}},
\end{array} \right.
\end{array}
\end{eqnarray}
where ${k_{\rm{F}}}$ is the Fermi wave number and $G_\sigma ^{\left( 0 \right)}\left( {{\mathbf{k}},\varepsilon } \right)$ the Green's function.

We introduce the following dimensionless quantities:
\begin{eqnarray}
\begin{array}{l}
\begin{array}{l}
Q = q/2{k_{\rm{F}}},\;K = k/2{k_{\rm{F}}};\;{k_{\rm{F}}} = {\left( {3{\pi ^2}n} \right)^{1/3}}\\
W = \hbar \omega /4{E_{\rm{F}}},\;E = \hbar \varepsilon /4{E_{\rm{F}}};\;{E_{\rm{F}}} = {\hbar ^2}k_{\rm{F}}^2/2{m^*}\\
\tilde G_\sigma ^{\left( 0 \right)} = 4{E_{\rm{F}}}G_\sigma ^{\left( 0 \right)}/\hbar \\
y = {m^*}{e^2}/{\hbar ^2}{k_{\rm{F}}},
\end{array}
\end{array}
\end{eqnarray}
where ${E_{\rm{F}}}$, $n$, and $v$ are the Fermi energy, the electron density, and volume of the system, respectively.

Substitution of these in Eq. (\ref{alfa}) gives
\begin{eqnarray}
\begin{array}{l}
{\alpha _0}\left( {{\bf{Q}},W} \right) =  - y\frac{{4\pi }}{{{Q^2}}}\sum\limits_\sigma  {\int {\frac{{{d^3}K}}{{{{\left( {2\pi } \right)}^3}}}} } \\
\int\limits_{ - \infty }^\infty  {\frac{{dE}}{{2\pi i}}\tilde G_\sigma ^{\left( 0 \right)}\left( {{\bf{K}},E} \right)\tilde G_\sigma ^{\left( 0 \right)}\left( {{\bf{K}} + {\bf{Q}},E + W} \right)} \\
 =  - y\frac{{4\pi }}{{{Q^2}}}\frac{1}{{v{{\left( {2{k_{\rm{F}}}} \right)}^3}}}\sum\limits_{{\bf{K}},\sigma } {} \\
\int\limits_{ - \infty }^\infty  {\frac{{dE}}{{2\pi i}}\tilde G_\sigma ^{\left( 0 \right)}\left( {{\bf{K}},E} \right)\tilde G_\sigma ^{\left( 0 \right)}\left( {{\bf{K}} + {\bf{Q}},E + W} \right)}. 
\end{array}
\end{eqnarray}

One can prove the following general relation:
\begin{equation}
\frac{\delta }{{\delta {n_{{\bf{P}},\sigma }}}}\sum\limits_{{\bf{K}},\sigma } {\int\limits_{ - \infty }^\infty  {\frac{{dE}}{{2\pi i}}\tilde G_\sigma ^{\left( 0 \right)}\left( {{\bf{K}},E} \right){F_{\sigma '}}\left( {{\bf{K}},E} \right)} }  = {F_\sigma }\left( {{\bf{P}},{P^2}} \right),
\end{equation}
which gives
\begin{eqnarray}
\begin{array}{r}
\frac{{\delta {\alpha _0}\left( {{\bf{Q}},W} \right)}}{{\delta {n_{{\bf{P}},\sigma }}}} =  - \frac{{4\pi }}{{{Q^2}}}\frac{1}{{v{{\left( {2{k_{\rm{F}}}} \right)}^3}}}\left[ {\tilde G_\sigma ^{\left( 0 \right)}\left( {{\bf{P}} + {\bf{Q}},{P^2} + W} \right)} \right.\\
 + \left. {\tilde G_\sigma ^{\left( 0 \right)}\left( {{\bf{P}} - {\bf{Q}},{P^2} - W} \right)} \right].
\end{array}
\end{eqnarray}

Now, the electron self-energy can be written on two alternative forms depending on from which of the two equivalent versions of the interaction energy in Eq. (\ref{Einter}) or Eq. (\ref{Einter2}) we start from. The two versions are
\begin{eqnarray}
\begin{array}{l}
\hbar {\Sigma _{{\bf{p}},\sigma }} =  - 16\pi y{E_{\rm{F}}}\int {\frac{{{d^3}Q}}{{{{\left( {2\pi } \right)}^3}}}\left[ {\int\limits_{ - \infty }^\infty  {\frac{{dW}}{{2\pi {\kern 1pt} i}}} \frac{{\tilde G_\sigma ^{\left( 0 \right)}\left( {{\bf{P}} + {\bf{Q}},{P^2} + W} \right)}}{{{Q^2}\varepsilon \left( {Q,W} \right)}}} \right.} \\
 + \left. {\frac{1}{{2{Q^2}}}} \right],
\end{array}
\label{SelfE1}
\end{eqnarray}
and
\begin{eqnarray}
\begin{array}{l}
\hbar {\Sigma _{{\bf{p}},\sigma }} =  - 16\pi y{E_{\rm{F}}}\int {\frac{{{d^3}Q}}{{{{\left( {2\pi } \right)}^3}}}\left\{ {\int\limits_{ - \infty }^\infty  {\frac{{dW}}{{2\pi {\kern 1pt} i}}} \left[ {\frac{{\tilde G_\sigma ^{\left( 0 \right)}\left( {{\bf{P}} + {\bf{Q}},{P^2} + W} \right)}}{{{Q^2}\varepsilon \left( {Q,W} \right)}}} \right.} \right.} \\
 - \left. {\left. {\frac{1}{{2{Q^2}}}\left( {\frac{1}{{W - \left[ {{{\left( {{\bf{P}} + {\bf{Q}}} \right)}^2} - {P^2}} \right] + i\eta }} - \frac{1}{{W + \left[ {{{\left( {{\bf{P}} + {\bf{Q}}} \right)}^2} - {P^2}} \right] - i\eta }}} \right)} \right]} \right\}.
\end{array}
\label{SelfE2}
\end{eqnarray}
In Eq. (\ref{SelfE1}) the contribution from the electron self-interaction is outside the $W$-integral but inside the $Q$-integral. In Eq. (\ref{SelfE2}) it is inside both.

We will now deform the integration path in the same way as we did to obtain Eq.(\ref{Einterimag}). In this case we will have poles from the Green's function inside the contour; if we
use Eq. (\ref{SelfE2}) we will also have poles from the second term.  we will get two contributions to the integral; one from the integration along the imaginary axis and one from
the residue contributions. These contributions are named the line- and residue-parts, respectively. The line-part is real-valued while the residue-part is complex-valued. Thus the
imaginary part of the self-energy comes entirely from the residue-part. One point to notice is that there is no unique separation in line- and residue-parts. This is obvious since in our two approaches the self-interaction term gives contributions to the residue-part in one of the approaches but not in the other.

Let us treat the line-part first. This contribution is
\begin{equation}
\begin{array}{l}
\hbar \Sigma _{{\bf{p}},\sigma }^{line} =  - 16\pi y{E_{\rm{F}}}\int {\frac{{{d^3}Q}}{{{{\left( {2\pi } \right)}^3}}}\left[ {\int\limits_{ - \infty }^\infty  {\frac{{dW}}{{2\pi {\kern 1pt} }}} \frac{{\frac{1}{{{P^2} + iW - {{\left( {{\bf{P}} + {\bf{Q}}} \right)}^2}}}}}{{{Q^2}\varepsilon \left( {Q,iW} \right)}}} \right.} \\
 + \left. {\frac{1}{{2{Q^2}}}} \right],
\end{array}
\end{equation}
where we have let the self-interaction term be included in the line-part. We use spherical coordinates and perform the angular integrations in the momentum integral. Then we make the substitution $W \to WQ$ and end up with the final result,
\begin{equation}
\begin{array}{l}
\hbar \Sigma _{{\bf{p}},\sigma }^{line} = \frac{{y{E_{\rm{F}}}}}{{{\pi ^2}}}\int\limits_0^\infty  {dQ} \left\{ {\int\limits_0^\infty  {dW\frac{1}{{P\varepsilon \left( {Q,iWQ} \right)}} \times } } \right.\\
\left. {\ln \left| {\frac{{{{\left( {Q + 2P} \right)}^2} + {W^2}}}{{{{\left( {Q - 2P} \right)}^2} + {W^2}}}} \right| - 4\pi } \right\}.
\end{array}
\label{Selfline}
\end{equation}

The residue-part has contributions if the poles of the Green's function happen to be
inside the contour. This happens in two cases:

\subsection{\label{Reshole}$P < \left| {{\bf{P}} + {\bf{Q}}} \right| < 1/2$}
The residue contribution to the self-energy for a hole is
\begin{equation}
\begin{array}{l}
\hbar \Sigma _{{\bf{p}},\sigma }^{res} =  - \frac{{2y{E_{\rm{F}}}}}{{{\pi ^2}}}\int\limits_{P < \left| {{\bf{P}} + {\bf{Q}}} \right| < 1/2} {{d^3}Q\frac{1}{{{Q^2}\varepsilon \left( {Q,{{\left( {{\bf{P}} + {\bf{Q}}} \right)}^2} - {P^2}} \right)}}}.
\end{array}
\end{equation}

\begin{figure}
\includegraphics[angle = 0, scale = 0.7]{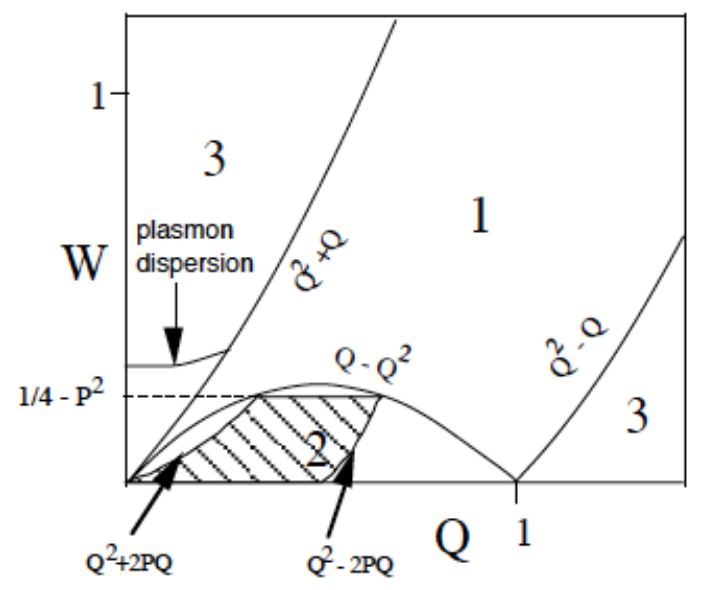}
\caption{\label{fig:4} The regions in the $WQ$-plane where the electron gas can gain energy and a hole in {\bf P} can loose energy. 
Regions $1$ and $2$ are the regions of electron-hole pair excitations. 
The curved line in the left part of region $3$ shows possible plasmon-excitations. 
The shaded area shows where a hole in {\bf P} can loose energy, $W$, and momentum, $Q$.}
\label{Fhole}
\end{figure}

This integral can be reformulated in a way that makes the physics more transparent. This is achieved with the substitution $W \to {\left( {{\mathbf{P}} + {\mathbf{Q}}} \right)^2} - {P^2}$. Then ${d^3}Q \to  - \left( {\pi /P} \right)QdQdW$ and
\begin{equation}
\hbar \Sigma _{{\mathbf{p}},\sigma }^{res} = 
 - \frac{{2y{E_{\text{F}}}}}{{\pi P}}
 {dWdQ\frac{1}{{Q\varepsilon \left( {Q,W} \right)}}},  
\label{Restransp}
\end{equation}
where the double integral is performed over the shaded area in Fig. \ref{Fhole}. 
The physics is the following: The hole in state ${\mathbf{P}}$ can fall upwards in energy. The energy and momentum is conserved via emission of electron-hole pairs; the conservation of energy only holds for actual transitions; these appear in the calculation of the imaginary part of the self-energy, which is closely connected to the life-time of the state. The shaded area shows which possible excitations (characterized by $(W,Q)$) are involved in the relaxation of the particular hole state. We realize that there is no possibility for plasmon excitations when the hole relaxes since the shaded area always stays within region 2.

\subsection{\label{Reselectron}$1/2 < \left| {{\bf{P}} + {\bf{Q}}} \right| <P $}
The residue contribution to the self-energy for an electron is
\begin{equation}
\hbar \Sigma _{{\mathbf{p}},\sigma }^{res} =  - \frac{{2y{E_{\text{F}}}}}{{{\pi ^2}}}\int\limits_{1/2 < \left| {{\mathbf{P}} + {\mathbf{Q}}} \right| < P} {{d^3}Q\frac{1}{{{Q^2}\varepsilon \left( {Q,{P^2} - {{\left( {{\mathbf{P}} + {\mathbf{Q}}} \right)}^2}} \right)}}}.
\end{equation}

\begin{figure}
\includegraphics[angle = 0, scale = 0.55]{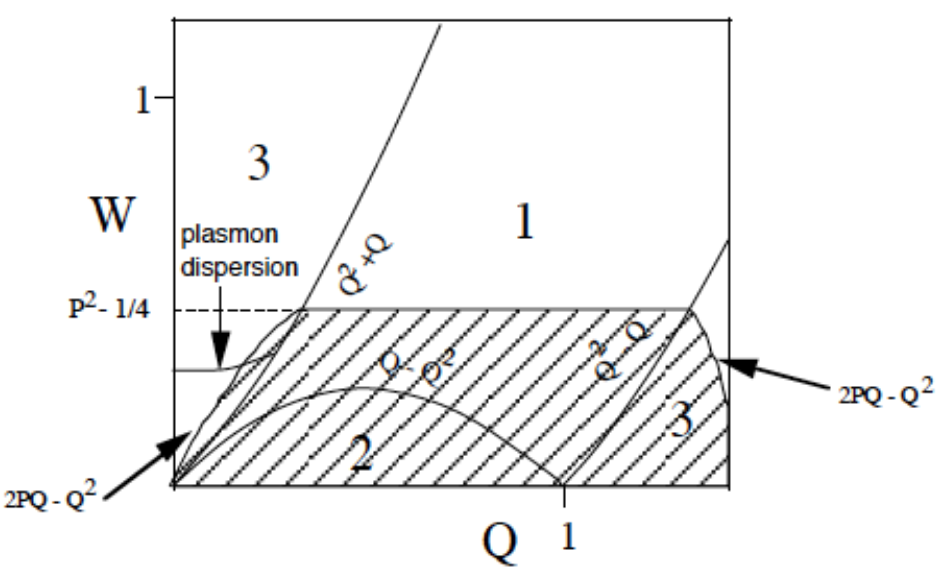}
\caption{\label{fig:5} The regions in the $WQ$-plane where the electron gas can gain energy and an electron in {\bf P} can loose energy. 
Regions $1$ and $2$ are the regions of electron-hole pair excitations. 
The curved line in the left part of region $3$ shows possible plasmon-excitations. 
The shaded area shows where an electron in {\bf P} can loose energy, $W$, and momentum, $Q$.}
\label{Felectron}
\end{figure}

The physics is more transparent by using again Eq. (\ref{Restransp}) where now the double integral is performed over the shaded area in Fig. \ref{Felectron}.

Here we see that the integration region extends outside the single-particle continuum. If the electron is high enough in energy, plasmons can be excited. We should notice that the
electron must have an energy somewhat larger than $\hbar {\omega _{pl}}$ above the Fermi energy to be able to excite plasmons; large-momentum plasmons are excited first (when the electron energy is gradually increased) and these have higher energy than the zero-momentum ones due to the dispersion of the plasmon line. This is also what is found in photo-emission experiments. The shaded area is the region that contains all possible energy and momentum pairs an electron in state p may give away when it relaxes towards equilibrium. It can only give away this combination if at the same time the rest of the system can absorb it. For the calculation of the imaginary part of the self-energy this limits the active part of the integration region to the union between the shaded area on the one hand and regions $1$ and $2$ and the plasmon line on the other. The integrand is basically the dynamical structure factor which is closely related to
the possible excitations of the system. With the above form of the integral we can actually separate out the various contributions to the decay of the electron.

Now we have completed the derivation of the electron self-energy from electron-electron interactions. However the integrals in the line-part is rather slowly converging. This
is improved if we use the approach with the artificial dielectric function. 

Appendix written by B.E.S.

\section{\label{B}Surface structure search by multiple scattering program  LEEDFIT}  
\subsection{\label{wB}Crystal potential with overlapping MT spheres}

The phase shift program EEASISSS is combined with the LEED multiple scattering program LEEDFIT. 
The phase shifts are calculated for the specific structure model defined in the LEED input file. 
The combination of the two programs performs the optimisation of the muffin-tin (MT) potential model 
in the LEED-I(V) analysis, and also to check the validity of the Sernelius's model for energy dependent 
exchange-correlation (XC) potential.

The LEED input file assigns a set of orbital phase shifts to each atom;
as orbitals are not present topic, it is understood that phase shifts are referred to atoms.
In order to limit the number of active phase shifts, the set of $N_\mathrm{ieq}$ symmetrically inequivalent 
atoms is grouped into $N_\mathrm{type}$ subsets, 
each of which contains a particular type of chemical element and its nearest neighbors. 
For example, in the case of  Cu(111)-$( 3\surd3\times\!\surd3 ) \mathrm{R30^\circ}$-TMB   
with one molecule C$_{24}$H$_{15}$S$_3$ per unit cell,
P3m1 symmetry reduces 24 C to 6 C that compose three groups of symmetrically different bonds, 
C-H, C-C, and C-S.
The assignment of each atom to the corresponding atom type is given in the LEEDFIT input.  
Each atom type comprises three parameters for the construction of the MT potential: 
minimum and maximum MT radii, $R_{\mathrm{min},i}$ and $R_{\mathrm{max},i}$, 
and a set of overlap parameters $S_i$, $i = 1,\ldots$,$N_{\mathrm{type}}$.
These items are called MT parameters, and structure parameters refer to the coordinates of atomic sites. 
In a LEED investigation there are in principle $3 \!\times\!N_\mathrm{type}$ variables to determine 
in addition to the structural parameters. 
In practice this is reduced in most cases to 1 parameter, the others are chosen automatically as discussed below.

This Supplemental Material considers the influence of the MT parameters on the LEED I(V) curves and 
shows the optimum choice of I(V) curves using Pendry's r-factor \cite{Pen80}.
A step-free MT potential with constant interstitial potential can be designed within a certain range of the MT parameters;
their optimum values are determined by the fit of calculated I(V) curves to experimental spectra.
In its original design the LEEDFIT program uses the layer doubling scheme, where the bulk backscattering 
matrices are calculated initially and reused in a refinement of the surface slab. 
The LEEDFIT program is here reorganized in such a way that MT parameters and structural parameters 
are optimizable simultaneously or separately. 
Each iteration of phase shifts gives rise to a new set of scattering matrices, and a special flow chart 
of LEED calculation is required in order to maintain bearable calculation times \cite{WM20}.

\begin{figure} 
  \includegraphics[angle = 0, scale = 0.35]{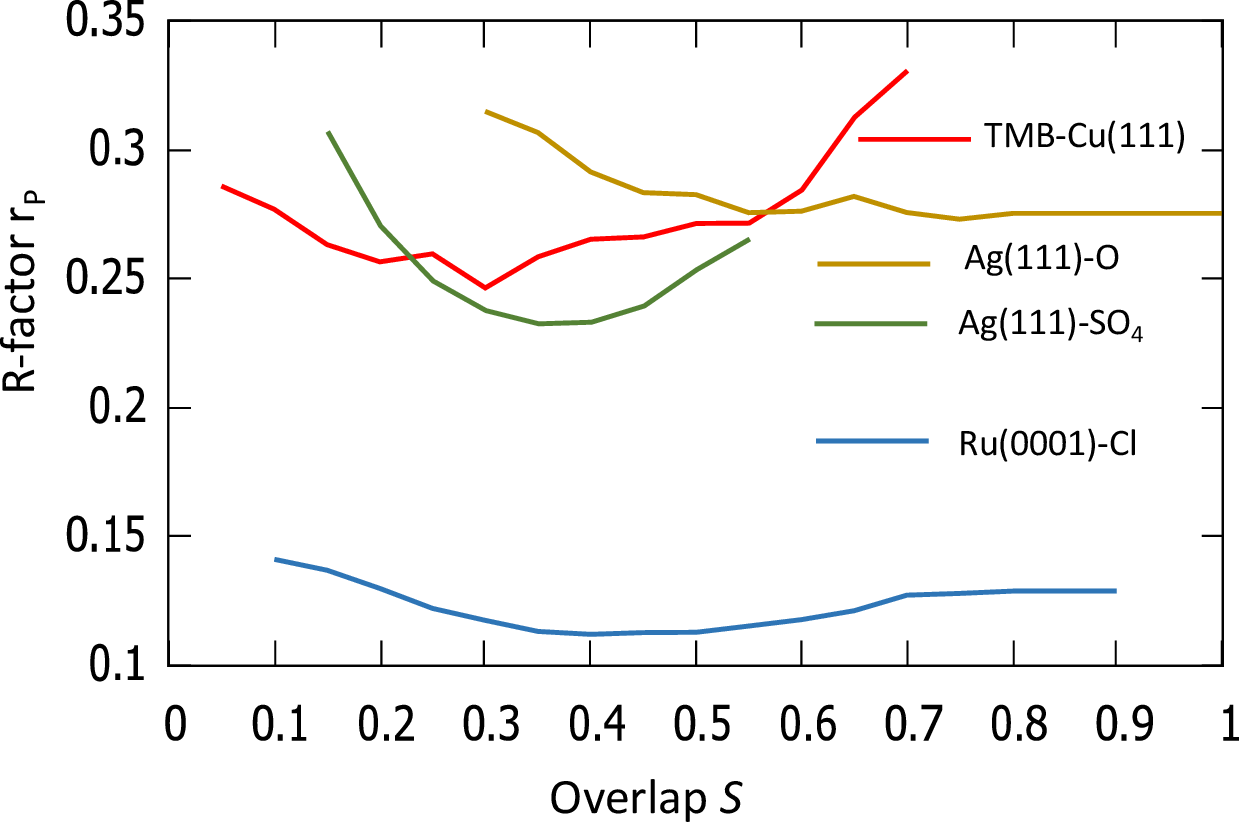}
  \caption{\label{fig:6} 
  R-factor as function of the overlap parameter for 
  Cu(111)-$( 3\surd3\times\!\surd3 ) \mathrm{R30^\circ}$-TMB \cite{Sir13}, 
  Ag(111)-$(4\!\times\!4 )$-O \cite{Rei07},  
  Ag(111)-$( 7\!\times\!\surd3 ) \mathrm{rect}$-SO$_4$ \cite{Wyr18}, 
  Ru(0001)-$( \surd3\!\times\!\surd3 ) \mathrm{R30^\circ}$-Cl \cite{Hof12}. 
  TMB stands for 1,3,5-tris(4-mercaptophenyl)-benzene with chemical formula C$_{24}$H$_{15}$S$_3$.}
 \end{figure}

Lower and upper limits of MT radii  are chosen automatically by LEEDFIT from the structure data. 
From interatomic distance $d_{\mathrm{NN},i}$ the minimal radius is set somewhat smaller than the 
midpoint radius, while the maximum radius is limited by the NN distance. 
The five surface structures investigated in Fig.~\ref{fig:6} indicate approximate limits for the MT radii: 
$R_{\mathrm{min},i} = 0.3\!\times\!d_{\mathrm{NN},i}$ and 
$R_{\mathrm{max},i} = 0.9\!\times\!d_{\mathrm{NN},i}$.
The overlap parameter $S_i$ is chosen variable between 0.0 to 1.0 and is fitted by r-factor minimization. 
$S_i$ can be chosen either as a single parameter common to all atom types or as separate overlap 
parameter for each different atom type.
At repeated instances during the structural r-factor minimization, LEEDFIT delivers iterated MT parameters
to EEASISSS that correspondingly generates iterated phase shifts.

\begin{figure} 
  \includegraphics[angle = 0, scale = 0.35]{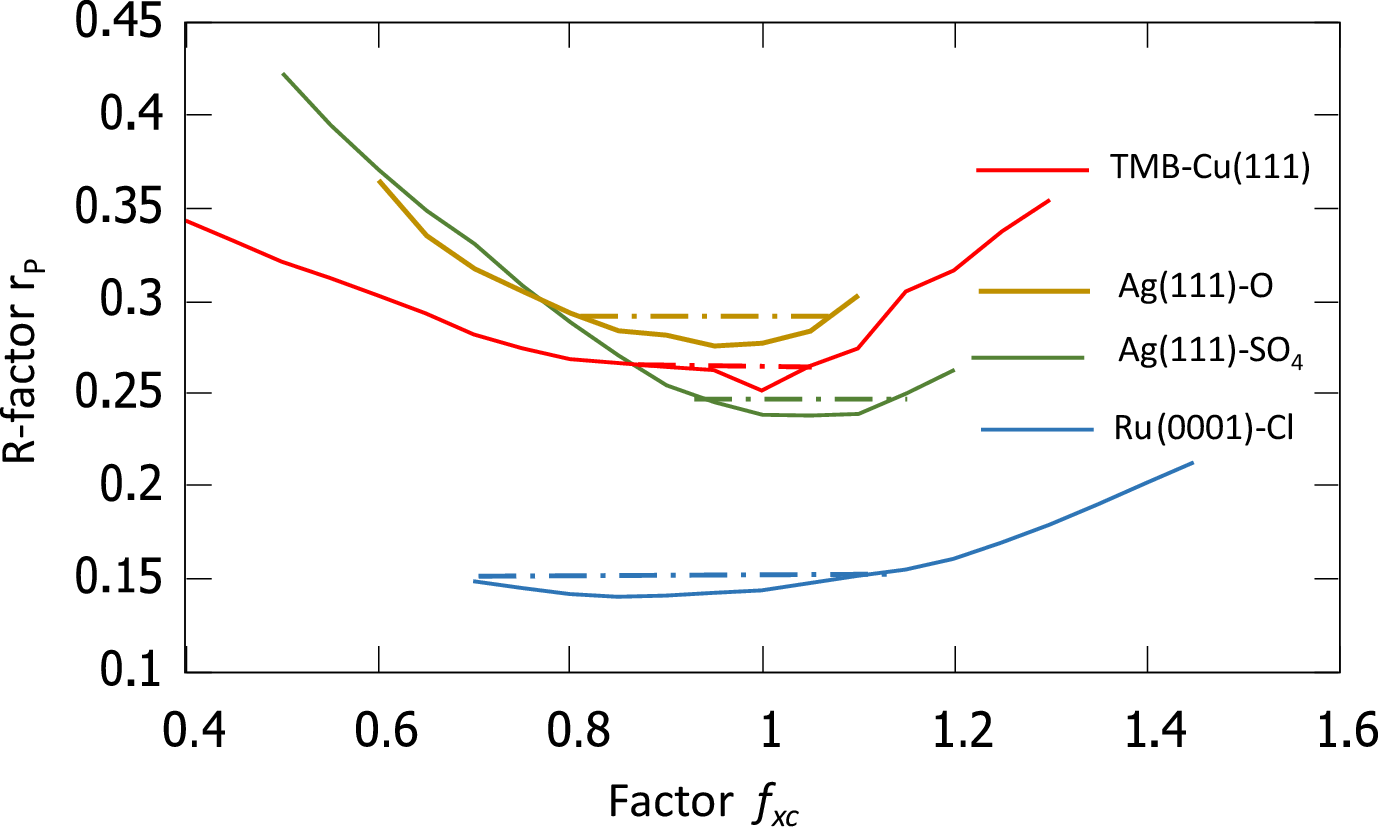}  
  \caption{\label{fig:7}  
  Influence of the self energy fitness factor $f_\mathrm{XC}$ on the r-factor for the 4 structures 
  investigated here. Labelling of the curves is same as in Fig.~\ref{fig:6}. 
  The range of each 
  $f_\mathrm{XC}$ curve extends between the end points of the dash-dotted RR line.
  $f_\mathrm{XC}$ value is measured at curve minimum and error bars from RR range: 
  $1.00\pm0.1$ (red), $0.94\pm0.1$ (brown), $1.06\pm0.1$ (green), and $0.89\pm0.2$ (blue). }
 \end{figure}

Figure \ref{fig:7} shows r-factors a function of overlap $S$ for four surface structures investigated in this work; 
the I(V) curves are available from best fit structure data; a common overlap turned out sufficient in each case.
The results for three of four cases show minimum r-factor in $S$ range $0.2 - 0.4$.
Ag(111)-$(4\!\times\!4 )$-O is an exception, where the minimum occurs at $S=0.75$.
Its r-factor remains constant at larger overlaps due to the maximum MT radius. 
The same situation appears for Ru(0001)-$( \surd3\!\times\!\surd3 ) \mathrm{R30^\circ}$-Cl at $S\geq 0.7$. 

We conclude that for initial structure search and refinement a value of $S = 0.3$ can be chosen for most atoms.
Changing MT parameters requires a subsequent refinement of structural parameters.
When a good model is found, an optimum overlap parameter is determined in a final refinement 
by a grid search or a fit. 
In the case of Ag(111)-$(4\!\times\!4 )$-O with minimum r-factor at $S=0.75$, the difference to the r-factor 
at $S=0.3$ is fairly small so that only minor structural differences are expected from a final refinement. 
The influence of the overlap parameter on the structural result has been found to be relatively small 
as discussed in the last section. 

\subsection{\label{wC} First LEED confirmation of electron self-energy potential }

To check whether the energy dependent exchange and correlation potential meets quantitatively the experimental data a factor $f_\mathrm{XC}$ was introduced to $V_{\mathrm{XC}0}$. The influence of this variable factor on the r-factor in the LEED-I(V) analysis is shown in Fig.~\ref{fig:8}. The error bar of the minimum r-factor defines a range for $f_\mathrm{XC}$. 
For the error bar the RR-factor defined by Pendry was used \cite{Pen80} which depends on the minimum r-factor and the total energy range of the I(V) curves.  A minimum around 1.0 was found in three cases. 
For Ru(0001)-$( \surd3\!\times\!\surd3 ) \mathrm{R30^\circ}$-Cl only few data could be measured due to the small unit cell. The minimum r-factor occurs   here at a slightly lower value but the r-factor curve is very flat and the value  $f_\mathrm{XC} = 1.0 $ lies in any case within the confidence interval. It appears sufficient to start the analysis with $f_\mathrm{XC}$ = 1.0 and to find the optimum with the final result. 

It should be noted that the experimental data include potential shifts from polarized adsorbates and further experimental errors due to measurement of the energy of the incident beam. 
Therefore, the parameters which are are fitted in the I(V) analyses describe the energy dependence of the inner potential, which is the difference between the external energy and the internal energy. These parameters deviate necessarily from the parameters for $V_\mathrm{XC}$ which are determined from Sernelius's model for the electron exchange and correlation potential. The experimental determined energy dependence of the inner potential does not 
directly proof the correctness of the model of $V_\mathrm{XC}$. 
The scattering inside the crystal takes place at energies given by the MT zero and $V_{\mathrm{XC}0}(E)$. 
Nevertheless, the finding that the minimum r-factor occurs at values $f_\mathrm{XC} \sim\! 1.0$ confirm the validity of Sernelius's model for the XC interaction.

\subsection{\label{wD}Influence of the upper and lower limit on the MT radii}

The lower and upper limits for the MT radii are automatically determined in the LEED program from the average NN distance for all atom types. An influence on the result is only found if $R_{\min,i}$ has been chosen 
to small or too large. In the phase shift program the MT radius for each atom type is chosen to be within the lower and upper limit. Mostly the limits  $R_{\min,i}$ = 0.3  and $R_{\max,i}$ = 0.9  are sufficient. 
The DE algorithm \cite{Sto97} in the phase shift program increases the actual MT radii by the overlap parameter for each atom type.
The criterion s to find a step-free MT potential. 
The parameter $R_{\min,i}$ provided from the LEED program sets therefore a lower limit for the overlap parameter.  There is a range for the overlap parameter $S_i$ where the actual MT radii remain within the limits and the r-factor remains at the minimum.
It can nevertheless occur in some structure models that lower values for $R_{\min,i}$ should be chosen.

(i) If the lower limit is too small the MT spheres do not touch, the phase shifts are not correct and the r-factor for the comparison with experimental data becomes worse. The lower limit has to be increased.

(ii) If the lower limit is too large two cases can occur. Either the MT radius of at least one atom type cannot be chosen such that a step-free potential can be found. Then the phase shift program fails.  Or the MT radius of at least one atom reaches the upper limit, the MT radii are not correctly chosen and the r-factor increases.

\subsection{\label{wE}Influence of the potential model on the structural parameters}

Each set of phase shifts calculated with slightly different MT radii generates slightly different scattering factors. 
  \begin{figure} 
  \includegraphics[angle = 0, scale = 0.35]{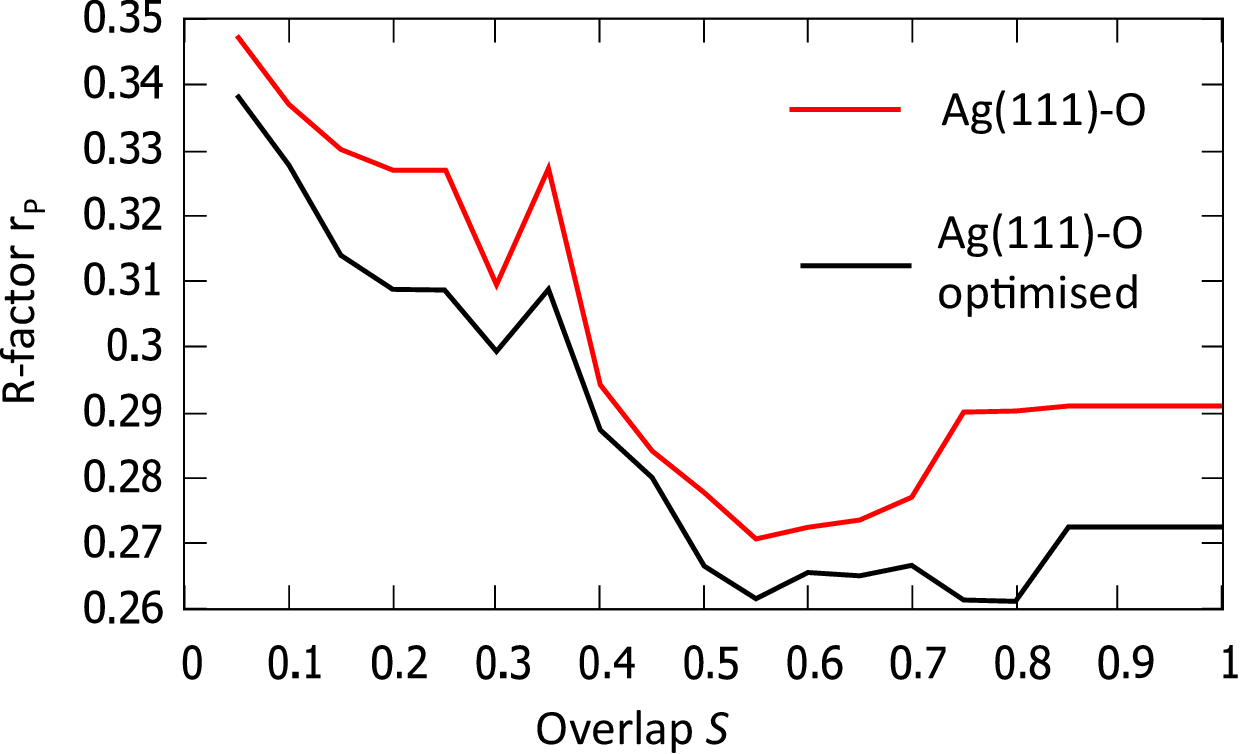} 
  \caption{\label{fig:8}  
  R-factors as function of the general overlap parameter. 
  The red curve is obtained by variation of the overlap  parameter $S$ and fix structural parameters from a previous 
  analysis with different phase shifts. 
  The black curve is obtained by variation  of $S$ and optimizing all structural parameters at each iteration. 
  The same minimum is found at $S = 0.55$ in both curves. 
  The black curve shows a minimum between 0.55 and 0.8. 
  The small increase between 0.55 and 0.8 is insignificant and demonstrates that a range of overlap exists where a 
  step-free potential can be obtained. }  
  \end{figure}
Different phases of the scattering factors lead to slightly different atomic positions if the structure is fitted. 
That means three sets of radial MT paramters add to the number of fit parameters in the structure analysis. 
Fortunately, it is not necessary to fit all parameters at once.

Usually, the phase shifts are not updated in each iteration of a fit or structure search procedure. If an optimum model is found the parameters for the phase shifts can be optimized separately keeping the structural parameters fix. In a subsequent final fit the structure parameter are found with optimized phase shifts. In three of the four cases investigated here where $20 - 50$ structural parameters were optimized we have found a noticeable improvement in the order of 0.02 up to 0.05 of Pendry{\rq}s r-factor. 

Figure \ref{fig:8} shows the r-factor as function of the general overlap parameter with structural parameters from a 
previous analysis and the same calculation with the final structural parameters. 
The two curves are mainly parallel and show the same minimum for the overlap parameter. 
This demonstrates that the MT potential can be separately optimized in a final step.  
and a simultaneous refinement of all parameters including the MT potential seems not necessary.  

The average change in the atomic positions is in the range of $0.02 - 0.05$ {\AA}.  In several cases a change of a single coordinate of an oxygen atom in the order of 0.1 {\AA} appeared. The structural changes remained in all cases within the error bars of the analysis.

Appendix written by W.M.

\bibliography{RefJ-210604} 
\end{document}